# Studies on tetrafluoropropene-CO$_2$ based gas mixtures for the Resistive Plate Chambers of the ALICE Muon IDentifier

**A. Ferretti** [1],[k] on behalf of the ALICE Collaboration and ECOGAS Collaboration:
**M. Abbrescia**[g], **G. Aielli**[b], **G. Alberghi**[c], **M. C. Arena**[q], **M. Barroso**[p], **L. Benussi**[d], **A. Bianchi**[k], **S. Bianco**[d], **D. Boscherini**[c], **A. Bruni**[c], **P. Camarri**[b], **R. Cardarelli**[a], **M. Corbetta**[n], **A. Di Ciaccio**[b], **L. Congedo**[g], **M. De Serio**[g], **L. Di Stante**[b], **P. Dupieux**[l], **J. Eysermans**[j], **M. Ferrini**[e], **M. Gagliardi**[k], **G. Galati**[g], **A. Gelmi**[g], **R. Guida**[n], **B. Joly**[l], **B. Liberti**[a], **B. Mandelli**[n], **S.P. Manen**[l], **L. Massa**[c], **L. Micheletti**[k], **L. Passamonti**[d], **A. Pastore**[r], **E. Pastori**[a], **D. Piccolo**[d], **D. Pierluigi**[d], **A. Polini**[c], **G. Proto**[a], **G. Pugliese**[g], **L. Quaglia**[k], **G. Rigoletti**[m,n], **M. Romano**[c], **A. Russo**[d], **A. Samalan**[i], **P. Salvini**[h], **R. Santonico**[b], **G. Saviano**[e], **S. Simone**[g], **L. Terlizzi**[k], **M. Tytgat**[i], **E. Vercellin**[k], **M. Verzeroli**[q] and **N. Zaganidis**[o]

[a] *INFN, Tor Vergata, Rome, Italy*
[b] *Dipartimento di Fisica di Roma Tor Vergata, Rome, Italy*
[c] *INFN, Bologna, Italy*
[d] *Laboratori Nazionali di Frascati dell'INFN, Italy*
[e] *Sapienza Universita di Roma, Dipartimento di Ingegneria Chimica Materiali Ambiente, Rome, Italy*
[f] *Laboratori Nazionali di Frascati dell'INFN, Italy*
[g] *Dipartimento di Fisica di Bari e sezione INFN di Bari, Italy*
[h] *Sezione INFN di Pavia, Italy*
[i] *Ghent University, Dept. of Physics and Astronomy, Proeftuinstraat 86, B-9000 Ghent, Belgium*
[j] *MIT, Cambridge, Massachusetts, USA*
[k] *Università di Torino and INFN, Sezione di Torino Via Giuria 1, 10125 Torino, Italy*
[l] *Clermont Université, Université Blaise Pascal, CNRS/IN2P3, Laboratoire de Physique Corpusculaire, BP 10448, F-63000 Clermont-Ferrand, France*
[m] *Université Claude Bernard Lyon I, Lyon, France*
[n] *CERN, Geneve, Switzerland*
[o] *Univ. Iberoamericana, Mexico City, Mexico*
[p] *Universidade do Estado do Rio de Janeiro*
[q] *Università degli Studi di Pavia, Italy*
[r] *Sezione INFN di Bari, Bari,*

*E-mail:* alessandro.ferretti@unito.it

ABSTRACT: Due to their simplicity and comparatively low cost Resistive Plate Chambers are gaseous detectors widely used in high-energy and cosmic rays physics when large detection areas are needed. However, the best gaseous mixtures are currently based on tetrafluoroethane, which has the undesirable characteristic of a large Global Warming Potential (GWP) of about 1400 and because of this, it is currently being phased out from industrial use. As a possible replacement, tetrafluoropropene (which has a GWP close to 1) has been taken into account.

---

[1] Speaker and corresponding author

Since tetrafluoropropene is more electronegative than tetrafluoroethane, it has to be diluted with gases with a lower attachment coefficient in order to maintain the operating voltage close to 10 kV. One of the main candidates for this role is carbon dioxide. In order to ascertain the feasibility and the performance of tetrafluoropropene-$CO_2$ based mixtures, an R&D program is being carried out in the ALICE collaboration, which employs an array of 72 Bakelite RPCs (Muon Identifier, MID) to identify muons. Different proportions of tetrafluoropropene and $CO_2$, with the addition of small quantities of isobutane and sulphur hexafluoride, have been tested with 50x50 $cm^2$ RPC prototypes with 2 mm wide gas gap and 2 mm thick Bakelite electrodes.

In the presentation, results from tests with cosmic rays will be presented, together with data concerning the current drawn by a RPC exposed to the gamma-ray flux of the Gamma Irradiation Facility (GIF) at CERN.

# Contents

## 1. ALICE MID RPCs and gas mixture GWP

The ALICE detector [1] is dedicated to the study of heavy-ion collisions at the CERN LHC and it is designed to study the physics of strongly interacting matter at extreme energy densities. ALICE is equipped with a Muon Spectrometer which tracks muons in the rapidity interval $2.5<\eta<4$. Muon identification in the spectrometer is performed by the Muon IDentifier (MID) [2] situated ~17 m from the Interaction Point downstream a 120 cm thick iron wall, covering an area of $5.5\times6.5$m. The MID is composed of an array of 72 Resistive Plate Chambers (RPCs) arranged in two stations, each with two detection planes. A typical RPC size is about $270\times70$ cm$^2$ and the maximum counting rate it has to withstand during physics runs is 100 Hz/cm$^2$.

ALICE MID RPCs are single gap (2 mm wide) and are made of resistive Bakelite electrodes 2 mm thick, with resistivity between $3\times10^9$ and $1\times10^{10}$ $\Omega\cdot$cm. The gas mixture is based on 89.7% tetrafluoroethane ($C_2H_2F_4$, R134a), which is a hydrofluorocarbon (HFC) gas. Isobutane (i-$C_4H_{10}$), and sulfur hexafluoride ($SF_6$) are added in percentages of 10% and 0.3%, respectively. The mixture is humidified (35-40% RH) in order to stabilize the electrodes resistivity. From Run 3 onwards, MID RPCs will work in avalanche mode: the signal induced on the readout strips will be processed by the ALICE FEERIC pre-amplified front-end discriminators, with a $Q_{induced}$ threshold of 130 fC. In these conditions, the effective working HVs are between 9.7 and 10 kV (HV values adjusted to 970 mbar of pressure and 20°C).

MID RPCs have been working smoothly since 2010 however, concerns about the gas mixture environmental impact arose. Both R-134a and $SF_6$ are Greenhouse Gases (GHG) with a Global Warming Potential (GWP$_{100yr}$) of respectively 1300 and 23500 [3]: the GWP of the full mixture amounts to 1351. Even with gas recirculation systems in place, during Run 1 ALICE RPCs accounted for 12.5% of the total CERN GHG emissions due to particle detectors [4]. If we add the GWP contribution due to similar gases emitted by RPCs installed in CMS and ATLAS experiments (where R134a constitutes 95% of the mixture) this fraction increases to 75%, more than 90% of which comes from R134a. Moreover, starting from January 2015 the European Union defined a set of environmental regulations under which HFC industrial use is progressively strongly restricted, driving up procurement costs. Therefore, the search for eco-friendly gas mixtures for RPCs operation plays a central role in the reduction of the CERN GHG emissions and costs. The ALICE MID group started to study the replacement of R134a in 2017 with cosmic rays. Since R134a is used in ALICE, ATLAS and CMS RPCs alike, in 2018 a common effort (ECOGAS collaboration) was started between RPC groups involved in these experiments, together with the CERN gas team. The SHIP collaboration joined the effort in 2019.

## 2. Cosmic rays tests

The goal of this R&D activity is to determine a low-GWP gas mixture that preserves detector performance, in particular, with respect to efficiency and streamer suppression, without impacting negatively on detector working parameters (i.e. the HV). To find a candidate to replace R134a we looked at HFO gases (Hydro-Fluoro-Olefin), which are similar to HFCs but with negligible GWP. Our choice fell on 1,3,3,3-Tetrafluoropropene ($C_3H_2F_4$, HFO-1234ze(E)), which has a similar chemical formula; it is not flammable at room temperature and features a GWP~1. One drawback is that electron capture is more dominant for tetrafluoropropene than in tetrafluoroethane: this means that a direct replacement would lead to a RPC HV working point (WP), well beyond 15 kV. In order to lower the HV, after preliminary studies we choose to add $CO_2$ to the $C_3H_2F_4$, $i-C_4H_{10}$ and $SF_6$ mix and then studied different percentages of these components.

The tests were performed in the INFN Torino laboratory, using one RPC with reduced dimensions (50 x 50 $cm^2$) and selecting cosmic rays with three plastic scintillators (total trigger area ~6 x 6 $cm^2$). The detector readout was the same as that used in the MID RPCs described above in Section 1. First, tests were done to understand the role of $C_3H_2F_4/CO_2$ ratio, with the addition of a fixed percentage of 10% $i-C_4H_{10}$ and 1% $SF_6$. Using 33.5% $C_3H_2F_4$ and 55.5% $CO_2$ we measured a ~1 kV efficiency shift towards higher voltages with respect to the standard ALICE mix with R134a. Raising $C_3H_2F_4$ to 44.5% and lowering $CO_2$ to the same value, the shift increase to ~2 kV (Fig. 1, left), while the streamer fraction at HV working point (i.e. the HV value when efficiency reaches 90%) was not significantly affected by the change. A similar behaviour was observed when varying the $C_3H_2F_4/i-C_4H_{10}$ ratio, confirming the higher electron affinity of tetrafluoropropene with respect to tetrafluoroethane.

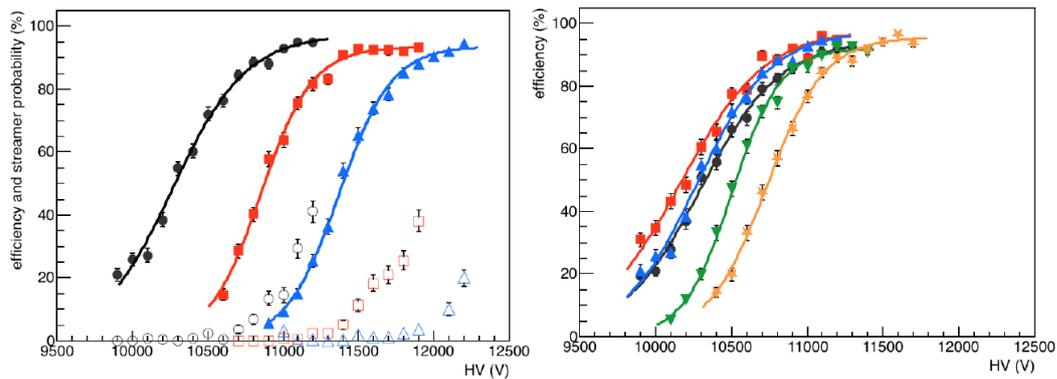

**Figure 1.** Left: efficiency and streamer comparison between mixtures with 10% $i-C_4H_{10}$, 1.0% $SF_6$ and 55.5% $CO_2$, 33.5% $C_3H_2F_4$ (black full), 50% $CO_2$, 39% $C_3H_2F_4$ (red full), 44.5% $CO_2$, 44.5% $C_3H_2F_4$ (blue full). The empty markers represent the respective streamer fractions. Right: Efficiency comparison between mixtures with 34.5% $C_3H_2F_4$, 1% $SF_6$ and 65.5% $CO_2$, 0% $i-C_4H_{10}$ (black), 60.5% $CO_2$, 5% $i-C_4H_{10}$ (red), 55.5% $CO_2$, 10% $i-C_4H_{10}$ (blue), 50.5% $CO_2$, 15% $i-C_4H_{10}$ (green), 45.5% $CO_2$, 20% $i-C_4H_{10}$ (orange)

We then studied the effect of varying the percentages of $CO_2$–$i$-$C_4H_{10}$ from 65.5%–0% to 44.5%-20% respectively, while keeping the amount of $C_3H_2F_4$ and $SF_6$ constant, respectively to 34.5% and 1%. It was observed that the HV working point varied erratically, decreasing by 100 V when raising the $i$-$C_4H_{10}$ fraction from 0% to 5% and then increasing up to +500 V when the isobutane reached 20% (Fig. 1, right), while the streamer fraction measured at HV working point remained basically unchanged between 5 and 10%.

After further tests of the $SF_6$ fraction, we ended up defining two promising mixture candidates, which are shown in Fig. 2 compared to the current ALICE mixture. The first one (50% $CO_2$, 39.7% $C_3H_2F_4$, 10% $i$-$C_4H_{10}$ and 0.3% $SF_6$) has a GWP of 72, but displays a HV shift of 1 kV and a higher streamer fraction with respect to the ALICE standard mixture. The second one differs basically only for the $SF_6$ percentage (1% instead of 0.3%): in this case, the HV shift is higher as well as the GWP (232, due to $SF_6$ addition), but the streamer fraction is closer to the ALICE standard.

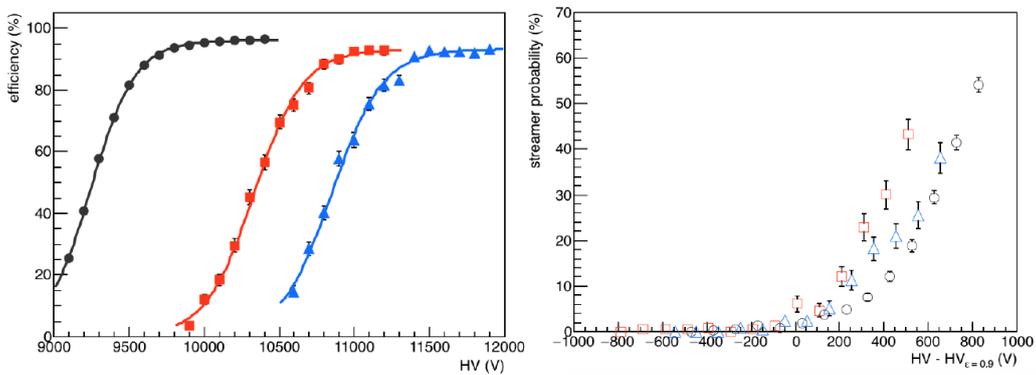

**Figure 2.** Left: efficiency comparison between the ALICE standard mixture (black), 50% $CO_2$, 39.7% $C_3H_2F_4$, 10% $i$-$C_4H_{10}$ and 0.3% $SF_6$ (red), 50% $CO_2$, 39% $C_3H_2F_4$, 10% $i$-$C_4H_{10}$ and 1% $SF_6$ (blue). Right: streamer probability as a function of the distance from the HV working point for the above mentioned mixtures (same color code applies)

## 3. Ageing tests under gamma irradiation

Finding suitable candidates is not enough to guarantee a successful replacement of the gas mixture until an ageing test is performed. Ageing processes in MID RPCs [6] are mostly due to deterioration of the inner electrodes surface smoothness which causes the formation of tips and ends up in an increase of the dark current. The deterioration depends on UV and chemical action, and it is strongly influenced by the chemical composition of the gas mixture. The exploration of ageing characteristics of HFO-based low-GWP mixtures is the common goal of the ECOGAS collaboration.

In order to check the ageing performance before replacement, active detectors are exposed to a high radiation dose in order to simulate many years of operations, keeping track of the performance over time. The ageing tests are being carried at the CERN Gamma Irradiation Facility (GIF++), equipped with a 14 TBq $^{137}$Cs source emitting 662 keV gamma rays. Lenticular lead filters can be used to modulate the radiation on the detectors under test.

Aging test have been performed since 2019, with two mixtures (ECO1 and ECO2), and work is still in progress. ECO1 recipe is 50% $CO_2$, 45% $C_3H_2F_4$, 4% $i$-$C_4H_{10}$ and 1% $SF_6$ while ECO2 is 60% $CO_2$, 35% $C_3H_2F_4$, 4% $i-C_4H_{10}$ and 1% $SF_6$. The main difference between these mixtures and the ALICE candidates is in the lower isobutane fraction, which allows treating the mixture as non-flammable from the safety point of view.

In the initial tests reported here, different RPC gas gaps were kept at a HV value close to the WP of the RPC and irradiated for some months (stability test). Every week a high voltage scan was performed in order to measure the current drawn by the detector when the gamma irradiation was off.

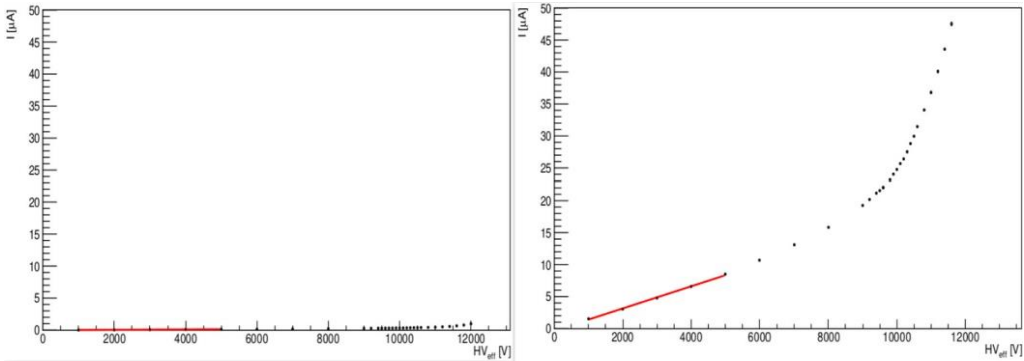

**Figure 3.** Left: dark current drawn by the ALICE 50×50 cm$^2$ prototype RPC as a function of the HV before the start of the irradiation test. Right: dark current drawn after the detector integrated a charge of 22 mC/cm$^2$ while flowed with ECO1 mixture.

Ageing tests are still ongoing and will continue in 2022. First results (Fig. 3) concerning the ALICE RPC show that after integrating a charge of 22 mC/cm$^2$ the dark current at the working point (estimated at 11.4 kV) is significantly higher with respect to the initial one. Moreover, the dark current at voltages lower than the electron multiplication threshold shows an ohmic behaviour which is not compatible with a simple mechanism of deterioration of the inner surface, because the tip effect cannot increase the electric field up to the point of provoking discharges even with voltages as low as 1 kV. This detector will soon be shipped back to the Torino laboratory in order to investigate the causes of this anomalous behaviour.